\documentclass[orivec]{llncs}
\usepackage[utf8]{inputenc}

\usepackage{tikz}
\usetikzlibrary{arrows}

\usepackage{amsfonts,amssymb,amstext}
\usepackage{wrapfig}
\usepackage{paralist}
\usepackage{ded}
\usepackage[hide]{ed}
\usepackage{basics}
\usepackage{local}
\usepackage[bookmarks,bookmarksnumbered]{hyperref}

\title{A Foundational View on Integration Problems
\thanks{preprint submitted to the Conference on Intelligent Computer Mathematics 2011, proceedings to appear in Springer LNCS}
}
\author{Florian Rabe$^1$ and Michael Kohlhase$^1$ and Claudio Sacerdoti Coen$^2$}
\institute{
 Computer Science, Jacobs University, Bremen (DE) \\
 \email{initial.lastname@jacobs-university.de}
 \and
 Department of Computer Science, University of Bologna (IT) \\
 \email{sacerdot@cs.unibo.it}
}

\begin{document}
\maketitle

\begin{abstract}
  The integration of reasoning and computation services across system and language
  boundaries is a challenging problem of computer science. In this paper, we use integration for the scenario where we have two systems that we integrate by moving problems and solutions between them. While this scenario is often approached from an engineering perspective, we take a foundational view.
  Based on the generic declarative language {\MMT}, we develop a theoretical framework for system integration using theories and partial theory morphisms. Because {\MMT} permits representations of the meta-logical foundations themselves, this includes integration across logics.
  We discuss safe and unsafe integration schemes and devise a general form of safe integration.
\end{abstract}

\ednote{FR: rewrote abstract (the reviewer didn't give details how to rewrite it, so I just rewrote it); it's rather short now, feel free to extend}

\section{Introduction}
  The aim of integrating Computer Algebra Systems (CAS) and Deduction Systems (DS) is
twofold: to bring the efficiency of CAS algorithms to DS (without sacrificing correctness)
and to bring the correctness assurance of the proof theoretic foundations of DS to CAS
computations (without sacrificing efficiency). In general, the integration of computation
and reasoning systems can be organized either by extending the internals of one system by
methods (data structure and algorithms) from the other, or by passing representations of
mathematical objects and system state between independent systems, thus delegating parts
of the computation to more efficient or secure platforms. We will deal with the latter
approach here, which again has two distinct sets of problems. The first addresses
engineering problems and revolves about communication protocol questions like shared
state, distributed garbage collection, and translating input syntaxes of the different
systems. The syntax questions have been studied extensively in the last decade and led to
universal content markup languages languages for mathematics like MathML and OpenMath to
organize communication. The second set of problems comes from the fact that passing
mathematical objects between systems can only be successful if their meaning is preserved
in the communication. This meaning is given via logical consequence in the logical system
together with the axioms and definitions of (or inscribed in) the respective systems.

We will address this in the current paper, starting from the observation that content level communication between mathematical systems, to be effective, cannot always respect logical consequence. On the other hand, there is the problem of trusting the communication itself, that boils down to studying the preservation of logical consequence. Surprisingly, this problem has not received in the literature the attention it deserves. Moreover, the problem of faithful safe communication, which preserves not only the consequence relation but also the intuitive meaning of a formal object, is not even always perceived as a structural problem of content level languages.

For example, people with a strong background in first order logic tend to assume that faithful and safe communication can always be achieved simply by strengthening the specifications; others believe that encoding logical theories is already sufficient for safe communication and do not appreciate that the main problem is just moved to faithfulness. Several people from the interactive theorem proving world have raised concerns about trusting CAS and solved the issue by re-checking the results or the traces of the computation (here called proof sketches). Sometimes this happens under the assumption that the computation is already correct and just needs to be re-checked, neglecting the interesting case when the proof sketch cannot be refined to a valid proof (or computation) without major patching (see~\cite{delahaye} for a special case).

In this paper, we first give a categorization of integration problems and solutions. Then we derive an integration framework by adding some key innovations to the {\MMT} language, a Module system for Mathematical Theories described in \cite{RK:mmt:10}. {\MMT} can be seen as a generalization of OpenMath and as a
formalized core of OMDoc. Of course, any specific integration task requires a substantial amount of work --- irrespective of the framework used. But our framework guides and structures this effort, and can implement all the generic aspects. In fact, current integration tasks typically involve setting up an ad-hoc framework for exactly that reason.

We sketch the {\MMT} framework first in Sect.~\ref{sec:integ:mmt}.
In Sect.~\ref{sec:integ:problem}, we analyze the integration problem for mathematical
systems from a formal position. Then we describe how integration can be realized our framework using partial {\MMT} theory morphisms in 
Sect.~\ref{sec:integ:solution}.
Finally, Sect.~\ref{sec:integ:related} discusses related work and Sect.~\ref{sec:integ:concl} concludes the paper.



\section{The MMT Language}\label{sec:integ:mmt}
  Agreeing on a common syntax like OpenMath is the first step towards system
integration. This already enables a number of structural services such as storage and
transport or editing and browsing that they do not depend on the semantics of the
processed expressions. But while we have a good solution for a joint syntax, it is
significantly harder to agree on a joint semantics. Fixing a semantics for a system
requires a foundational commitment that excludes systems based on other foundations. The
weakness of the (standard) OpenMath content dictionaries can be in part explained by this
problem: The only agreeable content dictionaries are those where any axioms (formal or
informal) are avoided that would exclude some foundations.

{\MMT} was designed to overcome this problem by placing it in between frameworks like
OpenMath and OMDoc on the one hand and logical frameworks like LF and CIC on the other
hand. The basic idea is that a system's foundation itself is represented as a content
dictionary. Thus, both meta and object language are represented uniformly as {\MMT}
\emph{theories}. Furthermore, \emph{theory morphisms} are employed to translate between
theories, which makes MMT expressive enough to represent translation between
meta-languages and thus to support cross-foundation integration.  As {\MMT} permits the
representation of logics as theories and internalizes the meta-relation between theories,
this provides the starting point to analyze the cross-foundation integration challenge
within a formal framework.

\paragraph{Syntax}
We will work with a very simple fragment of the {\MMT} language that suffices for our purposes, and refer to \cite{RK:mmt:10} for the full account. It is given by the following grammar where $[-]$ denotes optional parts and $T$, $v$, $c$, and $x$ are identifiers:
\begin{center}
\begin{tabular}{l@{\tb}l@{$\tb::=\tb$}l}
Theory graph   & $\TG$    & $\cdot \bnfalt \TG,\;\thdeclm{T}{[T]}{\theta} \bnfalt \TG,\;\vwdeclm{v}{T}{T}{v}{\sigma}$ \\
Theory body    & $\theta$ & $\cdot \bnfalt \theta,\;c\,[:\,O]\,[=\,O']$ \\
Morphism body  & $\sigma$ & $\cdot \bnfalt \sigma,\;\maps{c}{O}$ \\
Objects        & $O$      & {\OM} objects \\ 
Morphisms      & $\mu$    & $v \bnfalt \ident{T} \bnfalt \compose{\mu}{\mu}$ \\
Contexts       & $C$      & $x_1:O_1, \ldots, x_n:O_n$ \\
Substitutions  & $s$      & $x_1:=O_1, \ldots, x_n:=O_n$ \\
\end{tabular}
\end{center}
In particular, we omit the module system of {\MMT} that permits imports between theories.

$\thdeclm{T}{L}{\theta}$ declares a \emph{theory} $T$ with \emph{meta-theory} $L$ defined by the list $\theta$ of symbol declarations. The intuition of meta-theories is that $L$ is the meta-language that declares the foundational symbols used to type and define the symbol declarations in $\theta$.

All \emph{symbol declarations} in a theory body are of the form $\symdd{c}{O}{O'}$. This declares a new symbol $c$ where both the type $O$ and the definiens $O'$ are optional. If given, they must be $T$-objects, which are defined as follows. A symbol is called \emph{accessible} to $T$ if it is declared in $T$ or accessible to the meta-theory of $T$. An {\OM} object is called a $T$-\emph{object} if it only uses symbols that are accessible to $T$.

\begin{example}\label{ex:peano-cic}
Consider the natural numbers defined within the calculus of constructions (see~\cite{coq}). We represent this in {\MMT} using a theory $\CIC$ declaring untyped, undefined symbols such as $\cic{Type}$, $\lambda$ and $\to$. Then $\cic{Nat}$ is defined as a theory with meta-theory $\CIC$ giving symbol declarations such as $\cic{N}:\oms{\CIC}{\cic{Type}}$ or $\cic{succ}:\oma{\oms{\CIC}{\to}}{\oms{\cic{Nat}}{\cic{N}},\oms{\cic{Nat}}{\cic{N}}}$.
\end{example}

$S$-\emph{contexts} $C$ are lists of variable declarations $\ldots,x_i:O_i,\ldots$ for $S$-objects $O_i$. $S$-\emph{substitutions} $s$ for an $S$-context $C$ are lists of variable assignments $\ldots,x_i:=o_i,\ldots$. In an object $O$ in context $C$, exactly the variables in $C$ may occur freely; then for a substitution $s$ for $C$, we write $O[s]$ for the result of replacing every free occurrence of $x_i$ with $o_i$.

Relations between {\MMT} theories are expressed using theory morphisms. Given two theories $S$ and $T$, a \emph{theory morphism} from $S$ to $T$ is declared using $\vwdeclm{v}{S}{T}{l}{\sigma}$. Here $\sigma$ must contain one assignment $\maps{c}{O}$ for every symbol $c$ declared in the body of $S$, and for some $T$-objects $O$.
If $S$ and $T$ have meta-theories $L$ and $M$, then $v$ must also include a meta-morphism $l:L\to M$.

Every $\vwdeclm{v}{S}{T}{l}{\sigma}$ induces a \emph{homomorphic extension} $v(-)$ that maps $S$-objects to $T$-objects. $v(-)$ is defined by induction on the structure of {\OM} objects. The base case $v(c)$ for a symbol $c$ is defined as follows: If $c$ is accessible to the meta-theory of $S$, we put $v(c):=l(c)$; otherwise, we must have $\maps{c}{O}$ in $\sigma$, and we put $v(c):=O$. $v(-)$ also extends to contexts and substitutions in the obvious way.

By experimental evidence, all declarative languages for mathematics currently known can be represented faithfully in {\MMT}. In particular, {\MMT} uses the Curry-Howard representation \cite{curry,howard} of propositions as types and proofs as terms. Thus, an axiom named $a$ asserting $F$ is a special cases of a symbol $a$ of type $F$, and a theorem named $t$ asserting $F$ with proof $p$ is a special case of a symbol $t$ with type $F$ and definiens $p$. All inference rules needed to form $p$, are symbols declared in the meta-theory.

\paragraph{Semantics}
The use of meta-theories makes the logical foundation of a system part of an {\MMT} theory and makes the syntax of {\MMT} foundation-independent. The analogue for the semantics is more difficult to achieve: The central idea is that the semantics of {\MMT} is parametric in the semantics of the foundation.

To make this precise, we call a theory without a meta-theory \emph{foundational}. A \emph{foundation} for {\MMT} consists of a foundational theory $L$ and two judgments for typing and equality of objects:
\begin{compactitem}
	\item $\TG;C\vdash_T O:O'$ states that $O$ is a $T$-object over $C$ typed by the $T$-object $O'$,
	\item $\TG;C\vdash_T O=O'$ states the equality of two $T$-objects over $C$,
\end{compactitem}
defined for an arbitrary theory $T$ declared in $\TG$ with meta-theory $L$. In particular, {\MMT} does not distinguish terms, types, and values at higher universes --- all expressions are {\OM} objects with an arbitrary binary typing relation between them. We will omit $C$ when it is empty.

These judgments are similar to those used in almost all declarative languages, except that we do not commit to a particular inference system --- all rules are provided by the foundation and are transparent to {\MMT} except for the rules for the base cases of $T$-objects:
\[
\ibnc{\thdeclm{T}{L}{\theta} \minn \TG}{\symdd{c}{O}{O'}\minn\theta}{\TG\vdash_{T}c:O}{\TOtype}
\tb\tb
\ibnc{\thdeclm{T}{L}{\theta} \minn \TG}{\symdd{c}{O}{O'}\minn\theta}{\TG\vdash_{T}c=O'}{\TOdef}
\]
and accordingly if $O$ or $O'$ are omitted.
For example, adding the usual rules for the calculus of constructions yields a foundation for the foundational theory $\CIC$.

Given a foundation, {\MMT} defines (among others) the judgments
\begin{compactitem}
	\item $\TG\vdash \mu:S\to T$ states that $\mu$ is a theory morphism from $S$ to $T$,
	\item if $\TG\vdash\mu_i:S\to T$, then $\TG\vdash \mu_1= \mu_2$ states that $\vdash_T\mu_1(c)=\mu_2(c)$ for all symbols $c$ that are accessible to $S$,
	\item $\TG\vdash_S s:C$ states that $s$ is a well-typed for $C$, i.e., for every $x_i:=o_i$ in $s$ and $x_i:O_i$ in $C$, we have $\TG\vdash_S o_i:O_i$,
	\item $\TG\vdash \mathcal{G}$ states that $\mathcal{G}$ is a well-formed theory graph.
\end{compactitem}
In the sequel, we will omit $\TG$ if it is clear from the context.

The most important {\MMT} rule for our purposes is the rule that permits adding an assignment to a theory morphism: If $S$ contains a declaration $\symdd{c}{O_1}{O_2}$, then a theory morphism $\vwdeclm{v}{S}{T}{l}{\sigma}$ may contain an assignment $\maps{c}{O}$ only if $\vdash_T O:v(O_1)$ and $\vdash_T O=v(O_2)$. The according rule applies if $c$ has no type or no definiens. Of course, this means that assignments $\maps{c}{O}$ are redundant if $c$ has a definiens; but it is helpful to state the rule in this way to prepare for our definitions below.

Due to these rules, we obtain that if $\TG\vdash \mu:S\to T$ and $\vdash_S O:O'$ or $\vdash_S O=O'$, then $\vdash_T \mu(O):\mu(O')$ and $\vdash_T \mu(O)=\mu(O')$, respectively. Thus, typing and equality are preserved along theory morphisms.

Due to the Curry-Howard representation, this includes the preservation of provability: $\vdash_T p:F$ states that $p$ is a well-formed proof of $F$ in $T$. And if $S$ contains an axiom $a:F$, a morphism $\mu$ from $S$ to $T$ must map $a$ to a $T$-object of type $\mu(F)$, i.e., to a $T$-proof of $\mu(F)$. This yields the well-known intuition of a theory morphism. In particular, if $\mu$ is the identity on those symbols that do not represent axioms, then $\vdash \mu:S\to T$ implies that every $S$-theorem is an $T$-theorem.

{\MMT} is parametric in the particular choice of type system --- any type system can be used by giving the respective meta-theory. The type systems may themselves by defined in a further meta-theory. For example, many of our actual encodings are done with the logical framework LF \cite{lf} as the ultimate meta-theory. The flexibility to use {\MMT} with or without a logical framework that takes care of all typing aspects is a particular strength of {\MMT}.



\section{Integration Challenges}\label{sec:integ:problem}
  
In this section, we will develop some general intuitions about system integration and then
give precise definitions in {\MMT}. A particular strength of {\MMT} is that we can give
these precise definitions without committing to a particular foundational system and thus
without loss of generality.

The typical integration situation is that we have two systems $\sys_i$ for $i=1,2$ that
implement a shared specification $\Spec$. For example, these systems can be computer algebra
systems or (semi-)automated theorem provers. Our integration goal is to move problems and
results between $\sys_1$ and $\sys_2$.

\paragraph{Specifications and Systems}
Let us first assume a single system $\sys$ implementing $\Spec$, whose properties are
given by logical consequence relations $\Vdash_{\Spec}$ and $\Vdash_{\sys}$. We call
$\sys$ \emph{sound} if $\Vdash_{\sys} F$ implies $\Vdash_{\Spec} F$ for every formula $F$
in the language of $\Spec$. Conversely, we call $\sys$ \emph{complete} if $\Vdash_{\Spec}
F$ implies $\Vdash_{\sys} F$.

While these requirements seem quite natural at first, they are too strict for practical purposes. It is well-known that soundness fails for many CASs, which compute wrong results by not checking side conditions during simplification. Reasons for incompleteness can be theoretical --- e.g., when $\sys$ is a first-order prover and $\Spec$ a higher-order specification --- or practical --- e.g., due to resource limitations.

Moreover, soundness also fails in the case of underspecification: $\sys$ is usually much
stronger than $\Spec$ because it must commit to concrete definitions and implementations
for operations that are loosely specified in $\Spec$. A typical example is the
representation of undefined terms (see~\cite{DBLP:conf/cade/Farmer04} for a survey of
techniques). If $\Spec$ specifies the rational numbers using in particular $\forall
x.x\neq 0\impl x/x=1$, and $\sys$ defines $1/0=2/0= 0$, then $\sys$ is not sound because
$1/0=2/0$ is not a theorem of $\Spec$. 


We can define the above notions in {\MMT} as follows. A \emph{specification} $\Spec$ is an
{\MMT} theory; its meta-theory (if any) is called the \emph{specification language}. A
system implementing $\Spec$ consists of an {\MMT} theory $\sys$ and an {\MMT} theory
morphism $v:\Spec\to\sys$; the meta-theory of $\sys$ (if any) is called the
\emph{implementation language}.  With this definition and using the Curry-Howard
representation of {\MMT}, we can provide a deductive system for the consequence relations used above:
$\Vdash_{\Spec} F$ iff there is a $p$ such that $\vdash_{\Spec}p:F$; and accordingly for
$\Vdash_{\sys}$.

In the simplest case, the morphism $v$ is an inclusion, i.e., for every symbol in $\Spec$, $\sys$ contains a symbol of the same name. Using an arbitrary morphism $v$ provides more flexibility, for example, the theory of the natural numbers with addition and multiplication implements the specification of monoids in two different ways via two different morphisms.

\begin{example}\label{ex:peano-zf}
  We use a theory for second-order logic as the specification language; it declares
  symbols for $\forall$, $=$, etc. $\Spec=\Nat$ is a theory for the natural numbers; it
  declares symbols $\N$, $0$ and $succ$ as well as one symbol $\symdd{a}{F}{}$ for each Peano axiom $F$.

  For the implementation language, we use a theory $\ZF$ for ZF set theory; it has
  meta-theory first-order logic and declares symbols for $\zf{set}$, $\in$, $\es$, etc. Then we
  can implement the natural numbers in a theory $\sys=\zf{Nat}$ declaring, e.g., a symbol
  $\zf{0}$ defined as $\es$, a symbol $\zf{succ}$ defined such that $\zf{succ}(n)=n\cup\{n\}$, and prove
  one theorem $\symdd{a}{F}{p}$ in $\sys$ for each Peano axiom.  Note that $\zf{Nat}$ yields theorems about
  the natural numbers that cannot be expressed in $\Spec$, for example $\Vdash_{\ZF} 0\in
  1$.
  We obtain a morphism $\mu_1:\Nat\arr\zf{Nat}$ using $\maps{\N}{\zf{N}}$, $\maps{0}{\zf{0}}$ etc.
  
  Continuing Ex.~\ref{ex:peano-cic}, we obtain a different implementation $\mu_2:\Nat\to\cic{Nat}$ using $\maps{\N}{\cic{N}}$, $\maps{0}{\cic{0}}$ etc.
\end{example}

To capture practice in formal mathematics, we have to distinguish between the definitional
and the axiomatic method.  The \emph{axiomatic} method fixes a formal system $L$ and then
describes mathematical notions in $L$-theories $T$ using free symbols and axioms.  $T$ is
interpreted in models, which may or may not exist. This is common in model theoretical
logics, especially first-order logic, and in algebraic specification.  In {\MMT}, $T$ is
represented as a theory with meta-theory $L$ and with only undefined constants. In
Ex.~\ref{ex:peano-zf}, $L$ is second-order logic and $T$ is $\Spec$.

The \emph{definitional} method, on the other hand, fixes a formal system $L$ together with
a minimal theory $T_0$ and then describes mathematical notions using definitional
extensions $T$ of $T_0$. The properties of the notions defined in $T_0$ are derived as
theorems. The interpretation of $T$ is uniquely determined given a model of $T_0$. This is
common in proof theoretical logics, especially LCF-style proof assistants, and in set
theory. In Ex.~\ref{ex:peano-zf}, $L$ is first-order logic, $T_0$ is ZF, and $T$ is
$\sys$.

\paragraph{Types of Integration}
Let us now consider a specification $\Spec$ and two implementations $\mu_i:\Spec\to\sys_i$. To simplify the notation, we will write $\vdash$ and $\vdash_i$ instead of $\vdash_{\Spec}$ and $\vdash_{\sys_i}$.
We first describe different ways how to integrate $\sys_1$ and $\sys_2$ intuitively.

\emph{Borrowing} means to use $\sys_1$ to prove theorems in the language of $\sys_2$. Thus, the input to $\sys_1$ is a conjecture $F$ and the output is an expression $\vdash_1 p:F$. In general, since {\MMT} does not prescribe a calculus for proofs, the object $p$ can be a formal proof term, a certificate, proof sketch, or simply a yes/no answer.

\emph{Computation} means to reuse a $\sys_1$ computation in $\sys_2$. Thus, the input of $\sys_1$ is an expression $t$, and the output is a proof $p$ with an expression $t'$ such that $\vdash_1 p:t=t'$. To be useful, $t'$ should be simpler than $t$ in some way, e.g., maximally simplified or even normalized.

\emph{Querying} means answering a query in $\sys_1$ and transferring the results to $\sys_2$. This is similar to borrowing in that the input to $\sys_1$ is a formula $F$. However, now $F$ may contain free variables, and the output is not only a proof $p$ but also a substitution $s$ for the free variables such that $\vdash_1 p:F[s]$.

In all cases, a translation $I$ must be employed to translate the input from $\sys_1$ to
$\sys_2$.  Similarly, we need a translation $O$ in the opposite direction to translate the
output $t'$ and $s$ and (if available) $p$ from $\sys_2$ to $\sys_1$.


To define these integration types formally in {\MMT}, we first note that borrowing is a special case of querying if $F$ has no free variables. Similarly, computation is a special case of querying if $F$ has the form $t=X$ for a variable $X$ that does not occur in $t$. 

\begin{wrapfigure}{r}{2.8cm}
\vspace*{-2.5em}
\begin{tikzpicture}[yscale=.5,xscale=.8]
\node (S) at (0,0)
    {$\Spec$};
\node (L1) at (2,1.5)
    {$\sys_1$};
\node (L2) at (2,-1.5)
    {$\sys_2$};
\draw[-\arrowtip](S) -- node[above,near start] {$\mu_1$} (L1);
\draw[-\arrowtip](S) -- node[below,near start] {$\mu_2$} (L2);
\draw[-\arrowtip](L1) to[out=-60,in=60] node[right] {$O$} (L2);
\draw[-\arrowtip](L2) to[out=120,in=-120] node[right] {$I$} (L1);
\end{tikzpicture}
\vspace*{-3.5em}
\end{wrapfigure}

To define querying in {\MMT}, we assume a specification, two implementations, and
morphisms $I$ and $O$ as on the right. $I$ and $O$ must satisfy
$\compose{I}{O}=\ident{\sys_2}$, $\compose{\mu_1}{O}=\mu_2$, and
$\compose{\mu_2}{I}=\mu_1$. \ednote{CSC: given the first equation, the other two imply each other. Should we make this explicit?} Then we obtain the following \emph{general form of an
  integration problem}: Given an $\sys_2$-context $C$ and a query $C\vdash_2 ?:F$ (where
$?$ denotes the requested proof), find a substitution $\vdash_1 s:I(C)$ and a proof
$\vdash_1 p:I(F)[s]$. Then {\MMT} guarantees that $\vdash_2 O(p):F[O(s)]$ so that we
obtain $O(s)$ as the solution.
Moreover, only the existence of $O$ is necessary but not $O$ itself --- once a proof $p$ is found in $\sys_1$, the existence of $O$ ensures that $F$ is true in $\sys_2$, and it is not necessary to
translate $p$ to $\sys_2$.
\medskip

We call the above scenario \emph{safe bidirectional communication} between $\sys_1$ and $\sys_2$ because $I$ and $O$ are theory morphisms and thus guarantee that consequence and truth are preserved in both directions.
This scenario is often implicitly assumed by people coming from the first-order logic community. Indeed, if $\sys_1$ and $\sys_2$ are automatic or interactive theorem provers for first-order logic, then the logic of the two systems is the same and both $\sys_1$ and $\sys_2$ are equal to $\Spec$.

If we are only interested in \emph{safe directed communication}, i.e., transferring results from $\sys_1$ to $\sys_2$, then it is sufficient to require only $O$. Indeed, often $\mu_2$ is an inclusion, and the input parameters $C$ and $F$, which are technically $\sys_2$-objects, only use symbols from $\Spec$. Thus, they can be moved directly to $\Spec$ and $\sys_1$, and $I$ is not needed.

Similarly, the substitution $s$ can often be stated in terms of $\Spec$. In that case, $O$ is only needed to translate the proof $p$. If the proof translation is not feasible, $O$ may be omitted as well.
Then we speak of \emph{unsafe communication} because we do not have a guarantee that the communication of results is correct.
For example, let $\sys_1$ and $\sys_2$ be two CASs, that may compute wrong results by not
checking side conditions during simplification. Giving a theory morphism $O$ means that
the ``bugs'' of the system $\sys_1$ must be ``compatible'' with the ``bugs'' of $\sys_2$,
which is quite unlikely.

The above framework for safe communication via theory morphisms is particularly appropriate for the integration of axiomatic systems. However, if $\sys_1$ and $\sys_2$ employ different mathematical foundations or different variants of the same foundation, it can be difficult to establish the necessary theory morphisms. In {\MMT}, this means that $\sys_1$ and $\sys_2$ have different meta-theories so that $I$ and $O$ must include a meta-morphism. Therefore, unsafe communication is often used in practice, and even that can be difficult to implement.

Our framework is less appropriate if $\sys_1$ or $\sys_2$ are developed using the
definitional method. For example, consider Aczel's encoding of set theory in type
theory~\cite{aczel,werner-zfc}. Here $\sys_1=\zf{Nat}$ as in Ex.~\ref{ex:peano-zf}, and
$\sys_2=\cic{Nat}$ as in Ex.~\ref{ex:peano-cic}. Azcel's encoding provides the
needed meta-morphism $l:\ZF\to\CIC$ of $O$. But because $\zf{Nat}$ is definitional, we
already have $O=l$, and we have no freedom to define $O$ such that it maps the concepts of
$\zf{Nat}$ to their counterparts in $\cic{Nat}$.  Formally, in {\MMT}, this means that the
condition $\compose{\mu_1}{O}=\mu_2$ fails. Instead, we obtain two versions of the natural
numbers in CIC: a native one given by $\mu_2$ and the translation of $\zf{Nat}$ given by
$\compose{\mu_1}{O}$. Indeed, the latter must satisfy all $\ZF$-theorems including, e.g.,
$0\in 1$, which is not even a well-formed formula over $\cic{Nat}$. We speak of
\emph{faithful communication} if $\compose{\mu_1}{O}=\mu_2$ can be established even when
$\sys_1$ is definitional. This is not possible in {\MMT} without the extension we propose
below.

\section{A Framework for System Integration}\label{sec:integ:solution}
  In order to realize faithful communication within {\MMT}, we introduce \emph{partial theory morphisms} that can filter out those definitional details of $\sys_1$ that need not and cannot be mapped to $\sys_2$. We will develop this new concept in general in Sect.~\ref{sec:integ:partial} and then apply it to the integration problem in Sect.~\ref{sec:integ:design}.

\subsection{Partial Theory Morphisms in {\MMT}}\label{sec:integ:partial}

\paragraph{Syntax}
We extend the {\MMT} syntax with the production $O ::= \hid$.
The intended use of $\hid$ is to put assignments $\maps{c}{\hid}$ into the body of a morphism $\vwdeclm{v}{S}{T}{l}{\sigma}$ in order to make $v$ undefined at $c$. We say that $v$ \emph{filters} $c$. The homomorphic extension $v(-)$ remains unchanged and is still total: If $O$ contains filtered symbols, then $v(O)$ contains $\hid$ as a subobject. In that case, we say $v$ \emph{filters} $O$.

\paragraph{Semantics}
We refine the semantics as follows. A \emph{dependency cut} $D$ for an {\MMT} theory $T$
is a pair $(D_{type},D_{def})$ of two sets of symbols accessible to $T$.
Given such a dependency cut, we define \emph{dependency-aware judgments} 
$\TG\vdash_D O:O'$ and $\TG\vdash_D O=O'$ as follows. $\TG\vdash_D O:O'$ means that there is a derivation of $\TG\vdash_T O:O'$ that uses the rules $\TOtype$ and $\TOdef$ at most for the constants in $D_{type}$ and $D_{def}$, respectively. $\TG\vdash_D O=O'$ is defined accordingly.

In other words, if we have $\TG'\vdash_D O:O'$ and obtain $\TG'$ by changing the type of any constant not in $D_{type}$ or the definiens of any constant not in $D_{def}$, then we still have $\TG'\vdash_D O:O'$. Then a \emph{foundation} consists of a foundational theory $L$ together with dependency-aware judgments for typing and equality whenever $T$ has meta-theory $L$.

We make a crucial change to the {\MMT} rule for assignments in a theory morphism: If $S$ contains a declaration $\symdd{c}{O_1}{O_2}$, then a theory morphism $\vwdeclm{v}{S}{T}{l}{\sigma}$ may contain the assignment $\maps{c}{O}$ only if the following two conditions hold: (i) if $O_1$ is not filtered by $v$, then $\vdash_T O:v(O_1)$; (ii) if $O_2$ is not filtered by $v$, then $\vdash_T O=v(O_2)$. The according rule applies if $O_1$ or $O_2$ are omitted.

In \cite{RK:mmt:10}, a stricter condition is used. There, if $O_1$ or $O_2$ are filtered, then $c$ must be filtered as well. While this is a natural strictness condition for filtering, it is inappropriate for our use cases: For example, filtering all $L$-symbols would entail filtering all $\sys$-symbols.

Our weakened strictness condition is still strong enough to prove the central property of theory morphisms: If $\TG\vdash \mu:S\to T$ and $\vdash_D O:O'$ for some $D=(D_{type},D_{def})$ and $v$ does not filter $O$, $O'$, the type of a constant in $D_{type}$, or the definiens of a constant in $D_{def}$, then $\vdash_T \mu(O):\mu(O')$. The according result holds for the equality judgment.

Finally, we define the \emph{weak equality of morphisms} $\mu_i:S\to T$. We define $\vdash \mu_1\leq \mu_2$ in the same way as $\vdash \mu_1=\mu_2$ except that $\vdash_T\mu_1(c)=\mu_2(c)$ is only required if $c$ is not filtered by $\mu_1$. We say that $\vdash\eta:T\to S$ is a \emph{partial inverse} of $\mu:S\to T$ if $\vdash\compose{\mu}{\eta}=\ident{S}$ and $\vdash\compose{\eta}{\mu}\leq\ident{T}$.

\begin{example}\label{ex:integ:mmtnat}
  Consider the morphism $\mu_1:\Nat\to\zf{Nat}$ from Ex.~\ref{ex:peano-zf}. We build its partial inverse $\vwdeclm{\eta}{\zf{Nat}}{\Nat}{l}{\sigma}$. The meta-morphism $l$ filters all symbols of $\ZF$, e.g., $l(\es)=\hid$.
Then the symbol $\zf{\N}$ of $\zf{Nat}$ has filtered type and filtered definiens. Therefore, the conditions (i) and (ii) above are vacuous, and we use $\maps{\zf{\N}}{\N}$ in $\sigma$.
Then all remaining symbols of $\zf{Nat}$ (including the theorems) have filtered definiens but unfiltered types. For example, for $\symdd{\zf{0}}{\zf{N}}{\es}$ we have $\eta(\es)=\hid$ but $\eta(\zf{\N})=\N$. Therefore, condition (ii) is vacuous, and we map these symbols to their counterparts in $\Nat$, e.g., using $\maps{\zf{0}}{0}$ in $\sigma$. These assignments are type-preserving as required by condition (i) above, e.g., $\vdash_{\Nat}\eta(\zf{0}):\eta(\zf{\N})$.
\end{example}

\subsection{Integration via Partial Theory Morphisms}\label{sec:integ:design}

The following gives a typical application of our framework by safely and faithfully communicating proofs from a stronger to a weaker system:
\def\infinityax{a_\infty}
\def\infinityth{t_\infty}
\begin{example}
In \cite{IR:foundations:10}, we gave formalizations of Zermelo-Fraenkel ($ZFC$) set theory and Mizar's Tarski-Grothendieck set theory ($TG$) using the logical framework $LF$ as the common meta-theory.
$ZFC$ and $TG$ share the language of first-order set theory. But $TG$ is stronger than $ZFC$ because of Tarski's axiom, which implies, e.g, the sentence $I$ stating the existence of infinite sets (which is an axiom in $ZFC$) and large cardinals (which is unprovable in $ZFC$). For example, we have an axiom $\symdd{\infinityax}{I}{}$ in $ZFC$, and an axiom $\symdd{\mathit{tarski}}{T}{}$ and a theorem $\symdd{\infinityth}{I}{P}$ in $TG$. Many $TG$-theorems do not actually depend on this additional strength, but they do depend on $\infinityth$ and thus indirectly on $\mathit{tarski}$.

Using our framework, we can capture such a theorem as the case of a $TG$-theorem $\vdash_D p:F$ where $F$ is the theorem statement and $\infinityth\in D_{type}$ but $\infinityth\nin D_{def}$ and $\mathit{tarski}\nin D_{type}$.
We can give a partial theory morphism $\vwdeclm{v}{TG}{ZFC}{\ident{LF}}{\ldots, \maps\infinityth\infinityax,\ldots}$. Then $v$ does not filter $p$, and we obtain $\vdash_{ZFC}v(p):F$.
\end{example}

\begin{wrapfigure}r{2.3cm}\vspace*{-2.5em}
\begin{tikzpicture}[yscale=.6,xscale=.8]
\node (S) at (0,0)
    {$\Spec$};
\node (L1) at (2,1.5)
    {$\sys_1$};
\node (L2) at (2,-1.5)
    {$\sys_2$};
\draw[-\arrowtip](S) -- node[below,near end] {$\mu_1$} (L1);
\draw[-\arrowtip](S) -- node[above,near end] {$\mu_2$} (L2);
\draw[-\arrowtip,dashed](L1) to[out=180,in=90] node[above] {$\eta_1$} (S);
\draw[-\arrowtip,dashed](L2) to[out=180,in=-90] node[below] {$\eta_2$} (S);
\end{tikzpicture}\vspace*{-2.5em}
\end{wrapfigure}

Assume now that we have two implementations $\mu_i:\Spec\to \sys_i$ of $\Spec$ and partial
inverses $\eta_i$ of $\mu_i$, where $\sys_i$ has meta-theory $L_i$. This leads to the
diagram on the right where (dashed) edges are (partial) theory
morphisms. We can now obtain the translations $I\colon\sys_2\to\sys_1$ and
$O\colon\sys_1\to\sys_2$ as $I=\compose{\eta_2}{\mu_1}$ and
$O=\compose{\eta_1}{\mu_2}$. Note that $I$ and $O$ are partial inverses of each other.

As in Sect.~\ref{sec:integ:problem}, let $C\vdash_2 ?:F$ be a query in $\sys_2$. If $\eta_2$ does not filter any symbols in $C$ or $F$, we obtain the translated problem $I(C)\vdash_1 ?:I(F)$. Let us further assume that there is an $\sys_1$-substitution $\vdash_1 s:I(C)$ and a proof $\vdash_1 p:I(F)[s]$ such that $p$ and $s$ are not filtered by $\eta_1$. Because $I$ and $O$ are mutually inverse and morphism application preserves typing, we obtain the solution $\vdash_2 O(p):F[O(s)]$.

The condition that $\eta_2$ does not filter $C$ and $F$ is quite reasonable in practice: Otherwise, the meaning of the query would depend on implementation-specific details of $\sys_2$, and it is unlikely that $\sys_1$ should be able to find an answer anyway.
On the other hand, the morphism $\eta_1$ is more likely to filter the proof $p$. Moreover,
since the proof must be translated from $L_1$ to $L_2$ passing through $\Spec$, the latter
must include a proof system to allow translation of proofs. In practice this is rarely the
case, even if the consequence relation of $\Spec$ can be expressed as an inference
system. For example, large parts of mathematics or the {\OM} content dictionaries
implicitly (import) first-order logic and ZF set theory.

We outline two ways how to remedy this: We can \emph{communicate filtered proofs} or
change the morphisms to \emph{widen the filters} to let more proofs pass.

\paragraph{Communicating Filtered Proofs}
Firstly, if the proof rules of $\sys_1$ are filtered by $\eta_1$, what is received by $\sys_2$ after applying the output translation $O$ is a filtered proof, i.e., a proof object that contains the constant $\hid$. $\hid$ represents gaps in the proof that were lost in the translation.

In an extreme case, all applications of proof rules become $\hid$, and the only unfiltered parts of $O(p)$ are formulas that occurred as intermediate results during the proof. In that case, $O(p)$ is essentially a list of formulas $F_i$ (a proof sketch in the sense of~\cite{formal-proof-sketches}) such that $I(F_1) \wedge \ldots \wedge I(F_{i-1})\vdash_1 I(F_i)$ for $i=1,\ldots,n$. In order to refine $O(p)$ into a proof, we have to derive $\vdash_1 F_n$. Most of the time, it will be the case that $F_1,\ldots, F_{i-1} \vdash_2 F_i$ for all $i$, and the proof is obtained compositionally if $\sys_2$ can fill the gaps through automated reasoning. When this happens, the proof sketch is already a complete declarative proof.

\begin{example}
Let $\sys_1$ and $\sys_2$ be implementations of the rational numbers with different choices for division by zero. In $\sys_1$, division by zero yields a special value for undefined results, and operations on undefined values yield undefined results; then we have the $\sys_1$-theorem $t$ asserting $\forall a,b,c.a (b / c) \doteq (ab) / c$.
  In $\sys_2$, we have $n/0 \doteq 1$ and $n \% 0 \doteq n$; then we have the $\sys_2$-theorems $t_1,t_2,t_3$ asserting $\forall m,n.n \doteq (n/m) * m + n \% m$, $\forall m.m/m \doteq 1$, and $\forall m.m \% m \doteq 0$.
  \ednote{FR: Do we have a reference for a system that does division in this way?}
  
The choice in $\sys_2$ reduces the number of case analyses in basic proofs. But $t$ is not a theorem of $\sys_2$; instead, we only have a theorem $t'$ asserting $\forall a,b,c.c\not\doteq 0\Rightarrow a (b / c) \doteq (ab) / c$. On the other hand, $\sys_1$ is closer to common mathematics, but the $t_i$ are not theorems of $\sys_1$ because the side condition $m\neq 0$ is needed.
   
Hence, we do not have a total theory morphism $O:\sys_1\to\sys_2 $, but we can give a partial theory morphism $O$ that filters $t$. Now consider, for example, a proof $p$ over $\sys_1$ that instantiates $t$ with some values $A,B,C$. When translating $p$ to $\sys_2$, $t$ is filtered, but we can still communicate $p$, and $\sys_2$ can treat $O(p)$ as a proof sketch. Typically, $t$ is applied in a context where $C\not\doteq 0$ is known anyway so that $\sys_2$ can patch $O(p)$ by using $t'$ --- which can easily be found by automated reasoning.

Integration in the other direction works accordingly.
\end{example}

\paragraph{Widening the Filters}
\begin{wrapfigure}r{3.7cm}\vspace*{-2em}
\begin{tikzpicture}[xscale=.8,yscale=.7]
\node (S) at (-1.8,0)
    {$\Spec$};
\node (S') at (0,0)
    {$\Spec'$};
\node (L1) at (2,1.5)
    {$\sys_1$};
\node (L2) at (2,-1.5)
    {$\sys_2$};
\draw[\arrowtipmono-\arrowtip](S) -- (S');
\draw[-\arrowtip](S') -- node[below,near end] {$\mu'_1$} (L1);
\draw[-\arrowtip](S') -- node[above,near end] {$\mu'_2$} (L2);
\draw[-\arrowtip,dashed](L1) to[out=180,in=50] node[above] {$\eta'_1$} (S');
\draw[-\arrowtip,dashed](L2) to[out=180,in=-50] node[below] {$\eta'_2$} (S');
\end{tikzpicture}\vspace*{-2em}
\end{wrapfigure}

An alternative solution is to use additional knowledge about $\sys_1$ and $\sys_2$ to obtain a translation where $O(p)$ is not filtered. In particular, if $p$ is filtered completely, we can strengthen $\Spec$ by adding an inference system for the consequence relation of $\Spec$, thus obtaining $\Spec'$. Then we can extend the morphisms $\mu_i$ accordingly to $\mu'_i$, which amounts to proving that $\sys_i$ is a correct implementation of $\Spec$. Now $\eta_i$ can be extended as well so that its domain becomes bigger, i.e., the morphism $\eta_1$ and thus $O$ filter less proofs and become ``wider''.

Note that we are flexible in defining $\Spec'$ as required by the particular choices of $L_1$ and $L_2$. That way the official specification remains unchanged, and we can maximize the filters for every individual integration scenario.

\begin{example}[Continuing Ex.~\ref{ex:integ:mmtnat}]
A typical situation is that we have a theorem $F$ over $\zf{Nat}$ whose proof $p$ uses the Peano axioms and the rules of first-order logic but does not expand the definitions of the natural numbers. Moreover, if $\symdd{a}{A}{P}$ is a theorem in $\zf{Nat}$ that establishes one of the Peano axioms, then $p$ will refer to $a$, but will not expand the definition of $a$. Formally, we can describe this as $\vdash_D p:F$ where $\zf{0},a\in D_{type}$ but $\zf{0},a\nin D_{def}$.

We can form $\Spec'$ by extending $\Spec$ with proof rules for first-order logic and extend $\eta$ to $\eta'$ accordingly.
Since $\eta$ does not filter the types of $\zf{0}$ and $a$, we obtain a proof $\vdash_{\Spec}\eta'(p):\eta'(F)$ due to the type-preservation properties of our partial theory morphisms. Despite the partiality of $\eta'$, the correctness of this proof is guaranteed by the framework.
\end{example}

\ednote{reviewer: I would have appreciate even pen-and-paper instantiations of the concepts presented and not only of the notations. For instance, what would be the translations, filters etc. for published work of integration like:
\cite{hol_isahol,isahol_hol,hol_coq}}

Both ways to integrate systems are not new and have been used ad hoc in concrete integration approaches, see Sect.~\ref{sec:integ:related}. With our framework, we are able to capture them in a rigorous framework where their soundness can be studied formally.

\section{Related Work}\label{sec:integ:related}
  The MoWGLI project~\cite{mowgli-deliverables} introduced the concept of ``semantic
markup'' for specifications in the calculus of construction as distinct from the ``content
markup'' in OpenMath and OMDoc. This corresponds closely to the use of meta-theories in
{\MMT}: ``content markup'' corresponds to {\MMT} theories without meta-theory; and
``semantic markup'' corresponds to {\MMT} theories with meta-theory CIC.

A framework very similar to ours was given in \cite{trustablecommunication}. Our {\MMT} theories with meta-theory correspond to their biform theories, except that the latter adds algorithms. Our theory morphisms $I$ and $O$ correspond to their translations $\mathtt{export}$ and $\mathtt{import}$.
The key improvement of our framework over \cite{trustablecommunication} is that, using MMT's meta-theories, the involved logics and their consequence relations can be defined declaratively themselves so that a logic-independent implementation becomes possible. Similarly, using logic morphisms, it becomes possible to implement and verify the trustability conditions concisely.

Integration by borrowing is the typical scenario of integrating theorem provers and proof assistants. For example, Leo-II \cite{leo2} or the Sledgehammer tactic of Isabelle \cite{isahol_fol} ($\sys_2$) use first-order provers ($\sys_1$) to reason in higher-order logic. Here the input translation $I$ is partial inverse of the inclusion from first-order logic to higher-order logic. A total translation from modal logic to first-order logic is used in \cite{mspass}. In all cases, the safety is verified informally on the meta-level and no output translation $O$ in our sense is used. But Isabelle makes the communication safe by reconstructing a proof from the proof (sketch) returned by the prover.

The above systems are called on demand using an input translation $I$. Alternatively a
collection of $\sys_1$-proofs can be translated via an output translation $O$ for later
reuse in $\sys_2$; in that case no input translation $I$ is used at all. Examples are the
translations from Isabelle/HOL in HOL Light \cite{isahol_hol}, from HOL Light to
Isabelle/HOL \cite{hol_isahol}, from HOL Light to Coq \cite{hol_coq}, or from Isabelle/HOL
to Isabelle/ZF \cite{isahol_isazf}.  The translation from HOL to Isabelle/HOL is notable
because it permits faithful translations, e.g., the real numbers of HOL can be translated
to the real numbers of Isabelle/HOL, even though the two systems define them
differently. The safety of the translation is achieved by recording individual
$\sys_1$-proofs and replaying them in $\sys_2$. This was difficult to achieve even though
$\sys_1$ and $\sys_2$ are based on the same logic.

The translation given in~\cite{hol_coq} is the first faithful translation
from HOL proofs to CIC proofs. Since the two logics are different, in order to
obtain a total map the authors widen the filter by assuming additional
axioms on CIC (excluded middle and extensionality of functions). This
technique is not exploitable when the required axioms are inconsistent.
Moreover, the translation is suboptimal, since it uses excluded middle also
for proofs that are intuitionistic. To improve the solution, we could
use partial theory morphisms that map case analysis over boolean in HOL to
$\top$, and then use automation to avoid excluded middle in CIC when the
properties involved are all decidable.

In all above examples but~\cite{hol_coq}, the used translations are not verified within a logical framework. The Logosphere \cite{logosphere} project used the proof theoretical framework LF to provide statically verified logic translations that permit inherently safe communication. Here the dynamic verification of translated proofs becomes redundant. The most advanced such proof translation is one from HOL to Nuprl \cite{hol_nuprl}.

The theory of institutions \cite{institutions} provides a general model theoretical framework in which borrowing has been studied extensively \cite{borrowing} and implemented successfully \cite{hets}. Here the focus is on giving the morphism $I$ explicitly and using a model theoretical argument to establish the existence of some $O$; then communication is safe without explicitly translating proofs.

Integration by computation is the typical scenario for the integration of computer algebra systems, which is the main topic of the Calculemus series of conferences. For typical examples, see \cite{Maple-Mode} where the computation is performed by a CAS, and~\cite{hopr} where the computation is done by a term rewriting system. Communication is typically unsafe. Alternatively, safety can be achieved if the results of the CAS --- e.g., the factorization of a polynomial --- can be verified formally in a DS as done in \cite{harrison-cas} and \cite{sorge-trivialsymbolic}.

Typical applications of integration by querying are conjunctive query answering for a
description logic.  For example, in \cite{tptpsigmakee}, a first-order theorem prover is
used to answer queries about the SUMO ontology.

The communication of filtered proofs essentially leads to formal proof sketch in the sense of \cite{formal-proof-sketches}.
The idea of abstracting from a proof to a proof sketch corresponds to the assertion level proofs used in \cite{tramp} to integrate first-order provers.
The recording and replaying of proof steps in \cite{hol_isahol} and the reconstruction of proofs in Isabelle are also special cases of the communication of filtered proofs.

\section{Conclusion}\label{sec:integ:concl}
  In this paper we addressed the problem of preserving the semantics in protocol-based
integration of mathematical reasoning and computation systems. We analyzed the problem from
a foundational point of view and proposed a framework based on theory graphs, partial
theory morphisms, and explicit representations of meta-logics that allows to state solutions to the integration
problem.

The main contribution and novelty of the paper is that it paves the way towards a {\emph{theory of integration}}. Theoretically, via filtering, this theory could be able to combine faithfulness with static verification, which would be a major step towards the integration and merging of system libraries. Moreover, we believe it is practical because it requires only a simple extension of the {\MMT} framework, which already takes scalability issues very seriously~\cite{KRZ:mmttnt:10}.

We do not expect that our specific solution covers all integration problems that come up in practice. But we do expect that it will take a long time to exhaust the potential that our framework offers.

\ednote{future work: one-to-many translations}


\newcommand{\etalchar}[1]{$^{#1}$}

\end{document}